\documentclass[letter]{aa}  
\usepackage{graphicx}
\usepackage{txfonts}

\usepackage{longtable}
\usepackage{threeparttable}

\def\Jm{\ensuremath{\rm J\,m^{-2}}}
\def\gcc{\ensuremath{\rm g\,cm^{-3}}}
\def\intd{\,\mathrm{d}}

\begin{document} 

   \title{A dearth of small particles in debris disks}

\subtitle{An energy-constrained smallest fragment size}

   \author{S. Krijt
          \and
          M. Kama}
      
   \institute{Leiden Observatory, Leiden University, P.O. Box 9513, 2300 RA Leiden, The Netherlands\\ \email{krijt@strw.leidenuniv.nl}}

   \date{}

 
  \abstract
   {A prescription for the fragment size distribution resulting from dust grain collisions is essential when modelling a range of astrophysical systems, such as debris disks and planetary rings.}
   {While the slope of the fragment size distribution and the size of the largest fragment are well known, the behaviour of the distribution at the small size end is theoretically and experimentally poorly understood. This leads debris disk codes to generally assume a limit equal to, or below, the radiation blow-out size.}
   {We use energy conservation to analytically derive a lower boundary of the fragment size distribution for a range of collider mass ratios. Focussing on collisions between equal-sized bodies, we apply the method to debris disks.} 
   {For a given collider mass, the size of the smallest fragments is found to depend on collision velocity, material parameters, and the size of the largest fragment. We provide a physically motivated recipe for the calculation of the smallest fragment, which can be easily implemented in codes for modelling collisional systems. For plausible parameters, our results are consistent with the observed predominance of grains much larger than the blow-out size in Fomalhaut's main belt and in the \emph{Herschel} cold debris disks.}
   {}

   \keywords{ interplanetary medium -- minor planets, asteroids: general -- solid state: refractory   }

   \maketitle
%

\section{Introduction}

Fragmenting collisions are important in a range of astrophysical systems. While the slope of the fragment size distribution and the size of the largest fragment are well characterized and can be used confidently in models, the smallest fragment size is less well understood and is usually assumed to be constant for all collisions. We provide a framework for self-consistently calculating the smallest fragment size as a function of material and collision parameters (Section~\ref{sec:theory}), and discuss its implications for modelling debris disks (Section~ \ref{sec:applications}). 

Numerous experimental studies have looked at the fragment size distribution of destructive collisions, focussing on the slope of the power law(s), and on the size of the largest fragment \citep{davis1990,ryan1991,nakamura1991,ryan2000}. The smaller end of the size distribution has received considerably less attention; the smallest fragments are hard to count experimentally, and require a very high resolution to be captured in numerical simulations. Fragment distributions are therefore incomplete below sizes of $100$~$\mu$m, or masses below $10^{-3}$~gr \citep{fujiwara1977,takagi1984}. Molecular dynamics \citep[e.g.][]{dominiktielens1997} or smooth particle hydrodynamics \citep{geretshauser2010} simulations have limited resolution and tend to focus on the fragmentation threshold velocity rather than the smallest fragments.

\section{Minimum fragment size in a single collision}\label{sec:theory}

We consider collisions below the hypervelocity regime, i.e. the relative velocity of the colliders is much smaller than their internal sound speed, generally implying $v_{\rm rel}\lesssim1~$km/s. Based on experiments, we adopt the standard fragment size distribution

\begin{equation}
n(s) = C\cdot s^{-\alpha},
\end{equation}
with $3 < \alpha < 4$, and $C$ a coefficient we express below. While the mass is dominated by the largest particles, the surface area and thus the surface energy is dominated by the smallest fragments. As the creation of infinitely small fragments would require an infinite amount of energy, while the amount of kinetic energy available in a collision is finite, the power law must stop or flatten at some small fragment size. To the best of our knowledge, however, the regime of fragment sizes relevant for the analysis below has not yet been probed by available experimental data nor described theoretically in an astrophysical context. 

Assuming spherical fragments with sizes between $s_{\rm min}$ and $s_{\rm max} (\gg s_{\rm min})$, the total fragment mass and surface area are

\begin{equation}\label{eq:M_frag}
M_{\rm frag} = \frac{4\pi \rho C}{3(4-\alpha)} s_{\rm max}^{4-\alpha}, ~~~~~~
A_{\rm frag} = \frac{4\pi C}{\alpha-3}  s_{\rm min}^{3-\alpha}.
\end{equation}
For a collision between two bodies of size $s_0$ and mass $M_0=(4\pi /3)\rho s_0^3$, mass conservation implies $M_{\rm frag}=2M_0$, and thus 

\begin{equation}\label{eq:C}
C = 2(4-\alpha)s_0^3 s_{\rm max}^{\alpha-4}.
\end{equation}

The pre-collision kinetic energy is simply $U_{\rm K} = (1/2) \mu v_{\rm rel}^2$, where $\mu=M_0/2$ denotes the reduced mass. The difference in surface energy before and after the collision equals $\Delta U_{\rm S}=\gamma(A_{\rm frag} - 8\pi s_0^2)$, where $\gamma$ equals the surface energy per unit surface of the material. Assuming that only a fraction $\eta$ of the kinetic energy is used for creating new surface, we can combine Eqs.~\ref{eq:M_frag} and \ref{eq:C} to obtain a lower limit for the smallest fragment size. For the specific case of $\alpha=3.5$, this reduces to
\begin{equation}\label{eq:s_min3.5}
s_{\rm min} =  \left(\frac{24 \gamma s_0}{\eta \rho s_0 v_{\rm rel}^2 + 24\gamma}\right)^2 s_{\rm max}^{-1},
\end{equation}
and gives the size of the smallest fragments created in a collision at $v_{\rm rel}$, assuming $\alpha=3.5$ and $s_{\rm max} \gg s_{\rm min}$.

Instead of forming a fragment distribution, we imagine the limiting case in which the kinetic energy just suffices to split both colliders in half\footnote{One could imagine splitting only one of the colliders, or indeed chipping off only small parts of one of the collider bodies. Since less surface area is created, this would still be allowed at very low velocities. However, in that case the largest fragment is of the same size as $s_0$. We refrain from identifying this as fragmentation, and use the size derived in Eq.~\ref{eq:s_split} as the size below which fragmentation becomes inefficient.}, i.e. $\eta U_{\rm K} = 2\pi s_0^2 \gamma$. Solving for $s_0$, we obtain 
\begin{equation}\label{eq:s_split}
s_0^{\rm split} = \frac{3 \gamma}{\eta \rho v_{\rm rel}^2},
\end{equation}
which is the smallest particle that can be split in half. The smallest fragment is slightly smaller, but does not have a rigorously defined radius because we assume spherical particles. Equation~\ref{eq:s_split} is similar to the result of \citet{biermann1980} if $\eta=1$. 

The same limit can be explored using Eq.~\ref{eq:s_min3.5}, by forcing $s_{\rm min} \sim s_{\rm max} \sim 2^{-1/3}s_0$. This results in 
\begin{equation}
\label{eq:biermann}
s_{\rm min} \simeq \frac{5 \gamma}{\eta \rho v_{\rm rel}^2},
\end{equation}
which is very similar to Eq.~\ref{eq:s_split}. To summarise, in an energetic collision in which many fragments are created, the size of the smallest fragment is given by Eq.~\ref{eq:s_min3.5}. When the relative velocity is decreased, the fragment distribution becomes more and more discrete, until we reach the limit described by Eq.~\ref{eq:biermann}, in which particles can only just be split into two.

\begin{figure}[!ht]
\includegraphics[clip=,width=.95\linewidth]{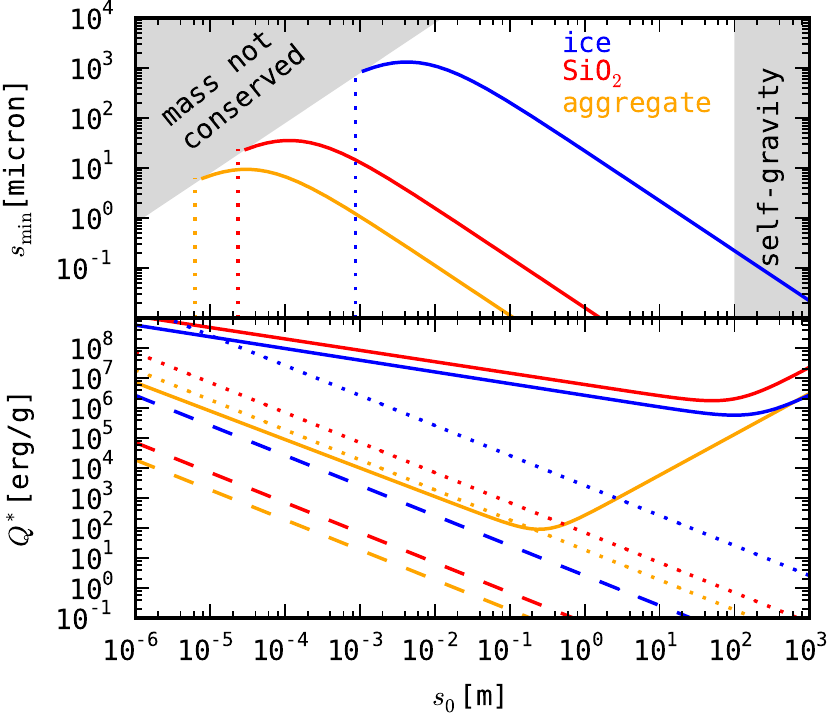}
\caption{\emph{Top:} Minimum fragment size $s_{\rm min}$ in a 20~m/s collision (Eq.~\ref{eq:s_min3.5}) for three materials, as a function of collider size $s_0$. We have assumed $\eta=10^{-2}$ and $s_{\rm max}=2^{-1/3}s_0$. The grey shaded areas are excluded because of mass conservation (left), and self-gravity (right). The dotted lines indicate the limit of Eq.~\ref{eq:biermann}. \emph{Bottom:} Critical energies versus size for equal-sized collisions. The energy needed to split a particle (e.g. Eq.~\ref{eq:biermann}) is shown for $\eta=10^{-3}$ (dotted) and $\eta=1$ (dashed). The solid curves correspond to catastrophic fragmentation of aggregates \citep{beitz2011}, and ice and basalt \citep{benz1999}, showing both the strength-dominated (small sizes) and self-gravity dominated regimes (large sizes).}
\label{fig:fig1}
\end{figure}

Figure \ref{fig:fig1} shows the minimum size from Eq.~\ref{eq:s_min3.5} as a function of collider size, assuming $v_{\rm rel}=20$~m/s, $\eta=10^{-2}$, and a maximum fragment that carries half of the initial collider mass. Gravity is important for bodies larger than $100$~m (see below). Smaller bodies are weaker, and can produce fragments down to the $s_{\rm min}$ indicated by the solid curve. For example, SiO$_2$ fragments smaller than a micron can, at this velocity, \emph{only} be formed by collisions of bodies larger than a few centimetres. The shaded region, top left, is forbidden, as there $M_{\rm frag} > 2M_0$ and mass is not conserved. Close to $s_{\rm min} \sim s_{0}$, the solid curves are non-linear as the pre- and post-collision surface areas become comparable.

\begin{table}
\caption{Material properties for silicate and ice used in this work. The values for the typical aggregate are explained in Appendix \ref{sec:A}.}
\label{tab:materials}      
\centering
\begin{tabular}{l c c}
\hline\hline
Material & $\rho$ (\gcc) & $\gamma$ (\Jm) \\
\hline
Silicate & 2.6 & 0.05 \\
Ice & 1.0 & 0.74 \\
Aggregate & $\sim10^{-1}$ & $\sim10^{-4}-10^{-3}$ \\
 \hline
\end{tabular}
\end{table}

It is interesting to compare Eq.~\ref{eq:biermann} with the traditional form of the catastrophic fragmentation threshold velocity in equal-sized collisions: $v_f^2 = 8Q^*$. The critical energy, $Q^*$, has units of $\rm{erg/g}$, and varies with particle mass. For small bodies, the strength is dominated by cohesion, and for large ones by gravity \citep{benz1999}. For solid bodies, this transition occurs around 100 metres in size. Values of $Q^*\sim10^7 \rm{~erg/g}$ are often taken as typical for asteroids, and experimentally obtained values for small grains (mm to cm sizes) can be several orders of magnitude smaller \citep{blummunch1993, beitz2011}. Figure \ref{fig:fig1} shows the critical energy for splitting predicted by Eq.~\ref{eq:biermann} as a function of size for the materials in Table \ref{tab:materials} assuming $\eta=10^{-3}$ and $\eta=1$. The solid lines indicate critical fragmentation energies for basalt and ice \citep{benz1999} and silicate aggregates \citep{beitz2011}. The critical fragmentation energies exceed the splitting energy, indicating that substantially more energy is required to destroy -- rather than split -- colliders. The values plotted for ice and basalt were obtained at collision velocities of 3~km/s, substantially higher than the velocities considered here, and $Q^*$ is known to depend on velocity \citep{Leinhardt2012}. While such a velocity dependence appears absent in the splitting energy, it might be implicitly included in $\eta$. In fact, $\eta$ is expected to vary with material and impact energy. We adopt a constant value of $\eta=10^{-2}$.

Appendix \ref{sec:B} investigates similar limits for colliders with different mass ratios, and shows that collisions with a mass ratio close to unity are the most effective at creating small fragments.

\section{Application to debris disks}\label{sec:applications}

Debris disks are leftovers of planet formation, and are usually described by a birth-ring of km-sized asteroid-like particles orbiting their parent star, together with a population of smaller bodies formed in a collisional cascade \citep[for a recent review, see][]{matthews2014}. A steady-state and scale-independent population of bodies will follow a size distribution given by a power law with $\alpha=3.5$ \citep{dohnanyi1969}. Some variation in $\alpha$ has been found in different simulations. \citet{PanSchlichting2012} find up to $\alpha=4$ for cohesion-dominated collisional particles, and up to $\alpha=3.26$ for gravity-dominated ones.

Models of debris disks most often assume a smallest fragment size equal to the blow-out size, $s_{\rm blow}$ \citep{WyattDent2002, Wyattetal2010}, or some constant, but arbitrary, $s_{\rm min} < s_{\rm blow}$ for all collisions \citep{Thebaultetal2003, Krivovetal2008}. The blow-out size corresponds to particles with $\beta=1/2$, where $\beta~=~1.15 Q_{\rm pr} (L_{\star}/L_{\odot}) (M_{\star}/M_{\odot})^{-1} (\rho/\mathrm{g~cm^{-3}})^{-1}(s/\mathrm{\mu m})^{-1}$ is the ratio of the radiation and gravitational force. Particles with $\beta>1/2$ are removed from the system by radiation pressure. Alternatively, \citet{Gasparetal2012a} calculate a collision-dependent $s_{\rm min}$ from mass conservation, but do not study the surface energy.

If, however, for any relevant collision Eq.~\ref{eq:s_min3.5} predicts $s_{\rm min}>s_{\rm blow}$, extrapolating the fragment size distribution down to these sizes is not justified. For example, starting from Eq.~2 of \citet{Krivovetal2008}, cm-sized bodies have $Q^*\simeq5\times10^6$ erg/g. In the Krivov et al. framework, a collision between two such bodies at 70~m/s will then result in fragmentation, as the kinetic energy ($\simeq 12$ J, assuming $\rho=2.35\rm{~g/cm^3}$) slightly exceeds the critical energy ($=2mQ^* \simeq 10$ J), and fragments will be created from a size comparable to the impactor \citep[Eq.~21 of][]{Krivovetal2006} down to the blow-out size. For this particular collision, Eq.~\ref{eq:s_min3.5} yields $s_{\rm min} \ll s_{\rm blow}$ for $\eta=0.1$ and $\gamma=0.1\rm{~J/m^2}$, but $s_{\rm min} \simeq 6\rm{~\mu m}$ for $\eta = 10^{-3}$. Thus, the difference between our results and the fragment sizes of Krivov et al. may be substantial, depending on the true value of $\eta$. We stress that our theory is valid for $3>\alpha>4$, and does not apply to models that use shallower power laws, for example Section 4.2 of \citet{Krivovetal2013}.

The importance of the limit given by Eq.~\ref{eq:s_min3.5} depends on the parameters, and can vary per individual collision, depending on the collision velocity and choice for $s_{\rm max}$. In the rest of this section, we explore in which cases this limit is most relevant.

In a debris disk, a particle of size $s_0$ is most likely formed in a collision between only slightly larger particles. In addition, we focus on collisions between equal-sized particles, as these are most efficient at forming small fragments (Appendix \ref{sec:B}). Therefore, we use Eq.~\ref{eq:biermann} as an indication for the lower limit of the particle size distribution. Quantitative comparisons require relative collision velocities, which for the largest bodies are often written in terms of the Keplerian orbital velocity at the corresponding distance from the central object. For bodies on orbits with identical semi-major axes, the relative velocity can then be written in terms of orbital eccentricity and inclination as $ f \equiv  v_{\rm rel} / v_{\rm K}  = (1.25 e^2 + i^2)^{1/2}$, with $v_{\rm K}=(GM_{\star} / a)^{1/2}$ the Keplerian orbital velocity \citep{WyattDent2002}. In a debris disk, a range of eccentricities and inclinations will be present. For a rough comparison, we use average quantities $\langle e \rangle$ and $\langle i \rangle$ to obtain typical collision velocities. In reality,  $\langle e \rangle$ and $\langle i \rangle$ are poorly constrained. Estimates range from $\langle e\rangle \sim \langle i\rangle \sim10^{-3}-10^{-1}$, depending on the level of stirring \citep{matthews2014}. 

\begin{figure}[!h]
\includegraphics[clip=,width=.95\linewidth]{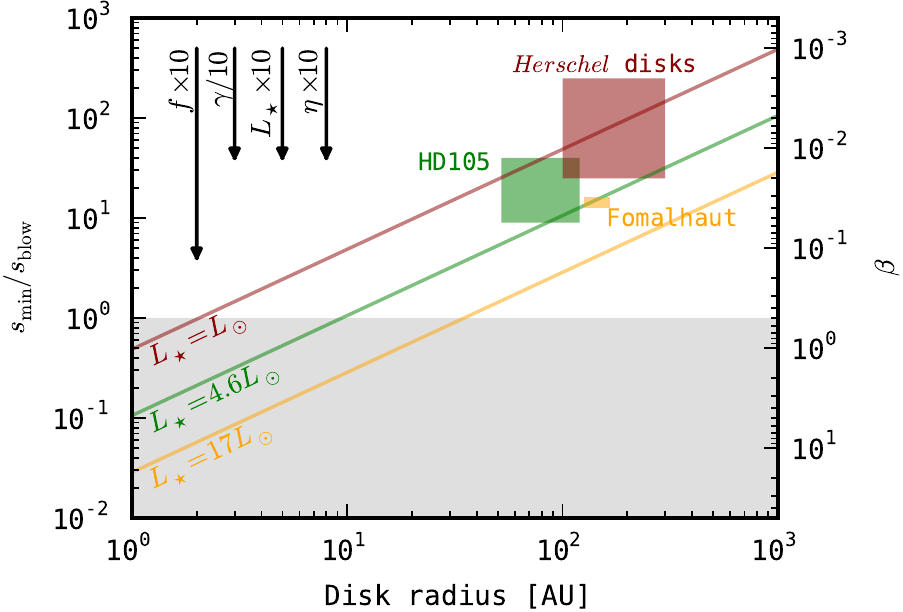}
\caption{Predicted $s_{\rm min}/s_{\rm blow}$ in debris disks, as a function of disk radius. Arrows indicate how the ratio changes with stellar luminosity, surface energy, $f$, and $\eta$. Coloured regions indicate observational constraints on $s_{\rm min}/s_{\rm blow}$ for various systems (see text), and the diagonal solid lines give our calculations for each system. We fix $\gamma=0.1\mathrm{~J~m}^{-2}$, $f=10^{-2}$, and $\eta=10^{-2}$ for this comparison.}
\label{fig:arrows}
\end{figure}

The ratio between the smallest grain size from Eq.~\ref{eq:biermann} and the blow-out size then becomes
\begin{equation}\label{eq:s_ratio}
\frac{s_{\rm min}}{s_{\rm blow}} = 2.4 \left( \frac{a}{5\mathrm{AU}} \right)  \left( \frac{L_{\star}}{L_{\odot}} \right)^{-1}  \left( \frac{f}{10^{-2}} \right)^{-2}  \left( \frac{\eta}{10^{-2}} \right)^{-1} \left( \frac{\gamma}{0.1~\mathrm{J~m^{-2}}} \right),
\end{equation}
where both the stellar mass and the material density drop out, and we assumed $Q_{\rm pr}=1$. Figure~\ref{fig:arrows} compares this ratio with observations of debris disks at large radii, where we predict the most pronounced effect. We have used a fixed $\gamma=0.1\mathrm{~J~m}^{-2}$, $f=10^{-2}$, and $\eta=10^{-2}$, and the arrows indicate the dependence of $s_{\rm min}/s_{\rm blow}$ on various parameters. For the main dust belt around \object{Fomalhaut}, \citet{Minetal2010} found the scattering properties to be consistent with predominantly $\sim100 \mu$m silicate grains \citep[$s_{\rm blow}=13\mu$m][]{Ackeetal2012}. The relative velocities in Fomalhaut are typically taken a factor of 10 higher \citep{WyattDent2002}. For \object{HD105}, \citet{Donaldsonetal2012a} derived $s_{\textrm{min}}=8.9 \mu$m ($s_{\rm blow}=0.5\mu$m) at orbital distances above $\sim50$~AU. Notably, very large grain sizes of $\sim100\mu$m ($s_{\rm blow}\lesssim1\mu$m) are inferred for the recently discovered ``\emph{Herschel} cold debris disks'' \citep{Krivovetal2013}, which are seen around F, G, and K type stars. \citeauthor{Krivovetal2013} were not able to model these systems with a collisional cascade reaching down to $s_{\rm min}=3\mu$m, and proposed that the large grains in these systems are primordial, unstirred material. Our calculations suggest that they can also be explained as the outcome of a collisional cascade. However, the model is highly degenerate, as material properties ($\eta$ and $\gamma$) and belt properties ($f$) are usually poorly known, and all have a large impact on $s_{\rm min}$.

In Figure~\ref{fig:arrows}, we assume constant and equal relative velocities for all particles. In reality, radiation pressure will also increase the eccentricities of small particles with $\beta \lesssim 0.5$. The enhanced eccentricity can be written as $e_{\beta}=\beta / (1-\beta)$. The relative velocity of such a radiation-influenced particle scales with its size as $v_{\rm rel} \propto \beta \propto s^{-1}$, while Eq.~\ref{eq:biermann} predicts the fragmentation velocity scales as $v_{\rm rel}\propto s_0^{-1/2}$. Hence, the relative velocity between the smallest particles increases \emph{faster} than the velocity needed for fragmentation. As a result, particles can reach arbitrarily small sizes in this regime. Particles smaller than $s_{\rm blow}$ are then removed on a short timescale. For a more detailed estimation, $\beta$ should be evaluated for each particular case, considering the optical properties of the material and the shape of the stellar spectrum.

A dearth of small grains in weakly stirred disks is also predicted by \citet{Thebault2008}, but the cause is not a limit on $s_{\rm min}$. In their scenario, $s_{\rm min}$ is fixed and the production rate of the smallest grains decreases with weaker stirring, while the destruction rate is determined by radiation forces and is unaffected by stirring. While the smallest grains present are always of blow-out size, their abundance is set by the balance between their creation and destruction (their Figure~7).

While the theory developed in this work predicts that the smallest particles that can fragment further can be quite large and sizes just below this will be depleted, some smaller particles will still be created as a result of erosive collisions, collisions between larger bodies, and collisions that occur above the average collision velocity. Detailed debris disk models implementing the surface energy constraint are needed to determine the resulting size distribution.

\section{Discussion}
Of the fundamental parameters in our model, the largest uncertainty affects $\eta$, the fraction of kinetic energy used for the creation of new surface. While information may be available about the kinetic and surface energy of the largest fragments, it is hard to quantify whether the remainder of the available energy went into surface creation, heat generation, or kinetic energy of the smallest fragments. Experimentally, studying $\eta$ is challenging, since it requires sensitive and complete measurements down to very small sizes. Once the functional form of $\eta$ is quantified by laboratory and numerical experiments, observations of $s_{\textrm{min}}$ in a system of interest may constrain $f$ and thus the local relative velocities.

During the preparation of this manuscript, we discovered that a similarly defined $s_{\textrm{min}}$ to the one we present has been explored in a more abstract framework, and without elaborating on applications, by \citet{Bashkirov1996}. We note that the lack of data on the size distribution of small collision fragments, as well as the fraction of kinetic and internal energy in the fragments already noted by \citeauthor{Bashkirov1996}, still prevails and we encourage further experiments to quantify these important parameters.

Thus far, we have focussed on equal-sized collisions. While collisions with a larger mass ratio might not lead to catastrophic fragmentation, cratering and erosion may still be important, and might be able to form small particles (Appendix \ref{sec:B}). Assuming a fixed relative velocity, we focus on a particle of size $s_1$. We define a mass loss rate $\dot{m}(s)$ for the larger particle, dependent on collider size $s$. Assuming a collision with a particle of size $s<s_1$ erodes a mass $\propto s^3$, and noting the collision timescale is proportional to the particle density and collision cross-section, we obtain for the \emph{total} mass loss rate $\dot{M} = \dot{m}(s)\intd s \propto s^{-3.5} s^3 (s+s_1)^2 \intd s$. If the collisional cross-section is dominated by $s_1$, we find $\dot{M} \propto s_1^2 s^{1/2} $, and thus the mass loss is dominated by the \emph{larger} bodies. When $s \sim s_1$, we obtain $\dot{M}\propto s^{5/2}$.

We have adopted a constant value of 50\% of the collider mass for the largest fragment. However, experiments show $s_{\rm max}$ can be substantially smaller as a function of material and impact velocity \citep[e.g.][]{davis1990,ryan1991}. Such results can easily be implemented in Eq.~\ref{eq:s_min3.5} (and Eqs.~\ref{eq:s_min_uneq} and \ref{eq:s_min_eros}) as necessary. Since $s_{\rm min} \propto s_{\rm max}^{-1}$, smaller sizes for the largest remnant will make the production of small particles even more difficult.

Other collisional systems where the proposed fragment size limit operates include planetary rings. Our calculations are consistent with the observed dominant grain sizes in the rings of Saturn, Jupiter, and Uranus. Because of additional relevant physics, such as tidal and electromagnetic effects, consistency does not directly imply the dominant grain size in all these rings is fragmentation-dominated.

The full implications of an energy-limited $s_{\textrm{min}}$ on systems such as debris disks and planetary rings can only be assessed with models tracking the full particle population with all relevant processes included. For example, if small particles cannot be destroyed in collisions, Poynting-Robertson (PR) drag will influence their orbits, and cause particles to drift towards the star on timescales of Gyrs in the outer parts of disks \citep{wyatt2005, lieshout2014}. Such modelling is outside the scope of the present paper.

\section{Conclusions}

We investigated the energetic constraints on the lower size limit in a distribution of collision fragments. A quantification of the lower limit of such size distributions is relevant for the modelling of debris disks and other astrophysical systems where collisional fragmentation is important.

   \begin{enumerate}
   \item{Based on surface energy constraints, we derive a parameterised recipe for the smallest fragment size in individual grain-grain collisions.}
   \item{The smallest size in a distribution of fragments from a two-particle collision, constrained by the collision energy, is given by Eq.~\ref{eq:s_min3.5}, and illustrated in Figure~\ref{fig:fig1}. For example, at 20 m/s, submicron silicate particles can only be effectively produced by centimeter-sized colliders.}
   \item{In the limit where the colliding bodies are split in half, the fragmentation threshold velocity is given by Eq.~\ref{eq:biermann}.}
   \item{While dedicated models are needed to reveal the full implications of the fragment size limit, Figure~\ref{fig:arrows} offers an indication of where the size distribution is expected to be influenced.}
   \item{In systems where the collision velocities are low, our theory may offer a natural explanation for a paucity of small grains in debris at large orbital distances, such as observed in Fomalhaut and the \emph{Herschel} cold debris disks (Figure~\ref{fig:arrows}).}
   \end{enumerate}

\begin{acknowledgements}
The authors wish to thank Carsten Dominik and Xander Tielens for comments and discussions. Dust studies at Leiden Observatory are supported through the Spinoza Premie of the Dutch science agency, NWO. Astrochemistry in Leiden is supported by NOVA, KNAW and EU A-ERC grant 291141 CHEMPLAN.
\end{acknowledgements}

\bibliographystyle{aa}
\bibliography{refs}

\Online

\begin{appendix}

\section{Applicability to aggregates}\label{sec:A}
For porous aggregates, the basic principles explored here are expected to hold, but some material properties have to be altered. First, aggregates have an internal filling factor $\phi = \rho_{\rm agg} / \rho$ that is $<1$, and might be as low as $10^{-4}$ in some extreme cases \citep{okuzumi2012,kataoka2013b}. Second, the 'effective' surface energy $\gamma_{\rm agg}$ will be smaller, since there is only limited contact between the aggregate's constituents to begin with. Assuming the parent bodies are built up of spherical monomers, the effective surface energy can be estimated as $\gamma_{\rm agg} \sim (a/R)^2 \phi^{2/3} \gamma$, where $a$ and $R$ denote the radius of the contact area shared by monomers, and the radius of the monomers themselves. The fraction $(a/R)$ depends on the size of the monomers and the material properties, but for 0.1-micron-sized monomers, $(a/R) \sim 0.1$ is reasonable.

For aggregates, $N$-body simulations have been performed with particles containing up to $10^6$ monomers, and values of $\eta$ range from close to unity \citep{dominiktielens1997,wada2009}, to several orders of magnitude less \citep{ringl2012b}, and depend on the employed contact model \citep{seizinger2013c}.

\section{Collisions with different mass ratios}\label{sec:B}
Here we extend the theory to collisions between 'targets' and 'projectiles' of arbitrary sizes $s_{\rm t} > s_{\rm p}$. Assuming the fragment distribution can be described as before, we can still use Eq.~\ref{eq:M_frag}, while the pre-collision kinetic energy now equals
\begin{equation}
U_{\rm K} = \frac{1}{2}\mu v_{\rm rel}^2 = \frac{1}{2} \frac{m_{\rm p} m_{\rm t}}{m_{\rm p}+m_{\rm t}} v_{\rm rel}^2.
\end{equation}
For a given collision velocity, we might then think of two cases, complete fragmentation when $s_{\rm p} \sim s_{\rm t}$, and erosion/cratering when $s_{\rm t} \gg s_{\rm p}$. 

\subsection{Catastrophic fragmentation}
Since both particles are destroyed completely we may set $M_{\rm frag} = m_{\rm p} + m_{\rm t}$. For simplicity we will assume in this section that the change in surface energy is dominated by the fragments $\Delta U_{\rm S}=\gamma A_{\rm frag}$. Using the same definition for $\eta$ as before and focussing on the $\alpha=3.5$ case, we obtain
\begin{equation}\label{eq:s_min_uneq}
s_{\rm min} = \left[ \frac{ 6 \gamma (s_{\rm p}^3+s_{\rm t}^3)^2}{\eta \rho v_{\rm rel}^2 (s_{\rm p}s_{\rm t})^3}\right]^2 s_{\rm max}^{-1}.
\end{equation}

\subsection{Erosion}
When the mass ratio becomes very large, it is no longer realistic to assume the target is completely disrupted. Rather, such collisions result in erosion, and the eroded mass is typically of the order of the mass of the projectile \citep{schrapler2011}. Thus, we write $M_{\rm frag} = \kappa m_{\rm p}$, with $\kappa$ of the order of unity. For large mass ratios $\mu \rightarrow m_{\rm p}$. Furthermore, assuming that the change in surface energy is dominated by the new fragments, $\Delta U_S = \gamma A_{\rm frag}$, we obtain
\begin{equation}\label{eq:s_min_eros}
s_{\rm min} = \left( \frac{ 6 \gamma \kappa}{\eta \rho v_{\rm rel}^2 }\right)^2 s_{\rm max}^{-1}.
\end{equation}

\begin{figure}[!ht]
\includegraphics[clip=,width=.95\linewidth]{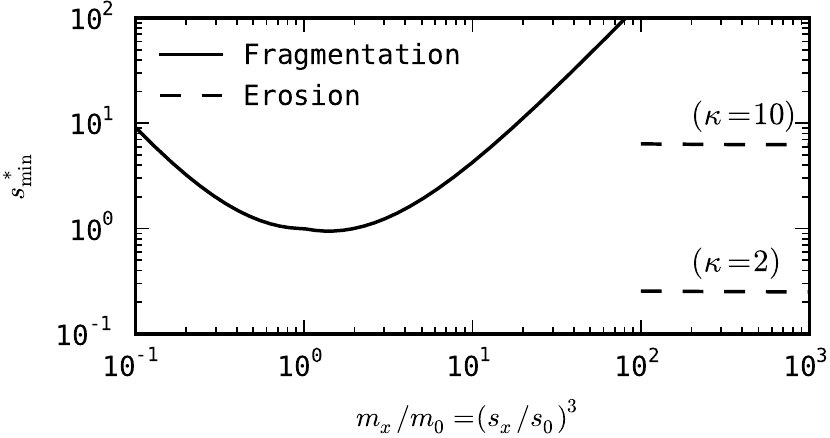}
\caption{Minimum fragment size resulting from destructive unequal collisions, assuming $s_{\rm min}\ll s_{\rm max}$, and relating the largest fragment size to the heavier collider. The y-axis has been normalized with the value for $s_{\rm min}$ in equal-mass collisions.}
\label{fig:fig4}
\end{figure}

Consider now a particle with a size $s_0$ close to the smaller end of the size distribution, colliding with particle of sizes $s_x$, ranging from slightly smaller to much larger than $s_0$. Figure \ref{fig:fig4} shows the minimum size of the fragments produced as a function of collider size $s_x$. The $y$-axis is normalized to the value obtained in equal-sized collisions (i.e. between two $s_0$ particles). For mass ratios below unity, $s_0$ acts as the target, and the mass of the largest fragment is assumed to equal $m_0/2$. For large mass ratios, $s_0$ acts as a projectile instead, and the largest fragment mass is set to $m_x/2$. Collisions at mass ratios above $10^2$ are assumed to be erosive \citep{seizinger2013c}, with the largest fragment equalling $m_0/2$. Since both the excavated mass, and the largest fragment mass, depend on the projectile, the curves in the erosive regime do not depend on the mass ratio directly. However, the size of the target does set an upper limit on $\kappa$.

\end{appendix}

\end{document}